\documentclass[12pt]{article}
\usepackage{latexsym, epsfig, graphics, amsmath,}
\numberwithin{equation}{section}
\newcommand{\be}{\begin{equation}}
\newcommand{\ee}{\end{equation}}
\newcommand{\bq}{\begin{eqnarray}}
\newcommand{\eq}{\end{eqnarray}}

\newcommand{\rhz}{\rho_{0}}
\newcommand{\lmz}{\lambda_{0}}
\newcommand{\eyz}{A_{0}(q)}

\newcommand{\mzs}{m_{0}^{2}}
\newcommand{\vts}{v(\tau,\sigma)}

\newcommand{\zbz}{\bar{Z}_{0}}
\newcommand{\zbi}{\bar{Z}_{I}}
\newcommand{\aaz}{\alpha_{0}}
\newcommand{\baz}{\beta_{0}}
\newcommand{\aao}{\alpha_{1}}
\newcommand{\bao}{\beta_{1}}

\newcommand{\cnb}{\bar{N}}
\newcommand{\xno}{x_{0}}
\begin{document}
\begin{titlepage}
\today          \hfill 
\begin{center}

%\hfill    LBNL-xxxxx \\
%          \hfill    UCB-PTH-xx/xx \\

\vskip .5in

{\large \bf More About QCD\,3 On The World Sheet}
\footnote{This work was supported 
 by the Director, Office of Science,
 Office of High Energy  
 of the U.S. Department of Energy under Contract No.
DE-AC02-05CH11231}.
\vskip .50in
K.Bardakci \footnote{Email: kbardakci@lbl.gov}

{\em Theoretical Physics Group\\
    Lawrence Berkeley National Laboratory\\
      University of California\\
    Berkeley, California 94720}
\end{center}

\vskip .5in

\begin{abstract}
  In this article, we extend the world sheet treatment of planar
 QCD in 1+2 dimensions from an earlier work. The starting point
 is a field theory that lives on the world sheet, parametrized
 by the light cone variables. In the present work, we generalize and
 extend the variational approach introduced earlier to get sharper
 results. An iterative solution to the variational equations leads to
 a solitonic ground state, and fluctuations around this ground
 state signals formation of a string on the world sheet. At high
 energies, the asymptotic limit of the string trajectory is linear,
 with calculable corrections at lower energies.

\end{abstract}
\end{titlepage}%THIS PAGE (PAGE ii) CONTAINS THE LBL DISCLAIMER

\newpage
\renewcommand{\thepage}{\arabic{page}}
\setcounter{page}{1}
%THIS IS PAGE 1 (INSERT TEXT OF REPORT HERE)
\section{Introduction}
\vskip 9pt

The present article is the continuation of a  previous
article [1]. The basic idea is to investigate planar $QCD\,3$,
using the world sheet methods and a variational ansatz developed in
[1]. $QCD\,3$ has been studied in the literature extensively
using various different approaches [2]. The world sheet formulation
we are going to use here was developed in [3,4,5].

The goal of the program is to sum the planar graphs of a field
theory on the world sheet prametrized by the light cone variables [6].
It was shown in [7] that this sum is reproduced by a two dimensional
field theory that lives on the world sheet. The challenge is to
find a manageable approximation scheme that captures the essence
of the model. The scheme used  in [1] was a
variational calculation, based on a simple ansatz. Here, we will use
the same type of ansatz; however, we will greatly enlarge the
parameter space of the ansatz by introducing a general variational
function $f(\sigma)$, to be determined by solving the variational
equations. For this purpose, we propose an iterative scheme based
on the expansion of $f(\sigma)$ in increasing powers of $\sigma$
around $\sigma=0$. This expansion leads to an asymptotic high enery
expansion of the fundamental theory. Based on this expansion, a
systematic method of solving the variational equations is developed.
In this paper, we work out only the first non-trivial term in the
expansion, and show that it results in a static solitonic solution.
This solution breaks translation invariance in the relative momentum
$q$, which has to be restored by introducing a collective coordinate,
to be identified with a string coordinate. In the rest of the paper,
the string picture based on this coordinate is developed. The main
result of the present work is the high energy asymptotic form of the
string trajectory: The leading term is linear, with non-leading
logarithmic and constant terms. There are additional non-leading
contributions, not calculated here, that vanish in the high energy
limit.

The following is a preview of the sections of this paper. Sections 1,
2 and 3 review the world sheet field theory that sums the planar
graphs in the light cone variables of $QCD\,3$. These sections are a
repetition of the corresponding sections in [1], and they are included
here for the convenience of the reader. In section 4, the variational
trial state is described. It depends on a function $A(q)$ of the
transverse momentum $q$,
and a function $f(\sigma)$ of the light cone coordinate
$\sigma$ mentioned above. The variational state is then constructed
by means of a recursion relation involving these functions.

In section 5, we derive and solve the equation obtained by setting
the variation of the Hamiltonian with respect to $A(q)$ equal
to zero. The solution depends on two constants $Z_{0}$ and $Z_{I}$,
which themselves depend on $f(\sigma)$. The ground state energy is
then expressed in terms of these constants, and it turns out to have
a linear divergence in the integral over the transverse momentum $q$.
This is due to translation invariance in this variable; the ground
state energy is proportional to the volume in the $q$ space. This type
of infinity is already known in the context of large $N$ matrix models
[8]. Here we
argue that the relevant finite quantity is the energy per unit volume.

In section 6, the recursion relations derived in section 4 are solved
by Fourier transform, and the results are expressed in terms of
$f(\sigma)$. At the end of the section, we write down the equation
obtained by setting the variation of the ground state energy with
respect to $f(\sigma)$ equal to zero. This is the fundamental
variational equation, whose solution will occupy the rest of the
paper.

In section 7, we write down the expansion of $f(\sigma)$ in powers
of $\sigma$, and work out the contribution of the first term in the
series to  the norm $N$ of the trial state. Since the result is a
summation of perturbation expansion, we argue that to get anything
different from perturbation, the denominator the geometric sum must
vanish. This condition determines the parameters of the first term
completely, and the variational function  is then 
the rest of the series, denoted by $\tilde{f}(\sigma)$. However,
the fixing of the first term introduces singularities in the auxiliary
functions in a certain parameter $s$. In the rest of the section,
we work out the dependence of these functions on $\tilde{f}$ and
on $s$. Later, we will show that these singularities cancel out from
the quantities of interest.

The variational equations for $\tilde{f}(\sigma)$ are still quite
formidable, and in this article, we will only solve for the first
term in the series for $\tilde{f}$. In section 8, the variational
equations for  the two constants $\bao$ and $x$, which parametrize
$\tilde{f}$, are derived and solved. $x$ is numerically fixed, and
$\bao$ turns out to be arbitrary. These are then the parameters of the
field configuration that solves the variational equations in the
leading approximation.

As was pointed out in [1], this configuration breaks translation
invariance in $q$. To restore this invariance, we introduce a
collective coordinate $v(\tau\,\sigma)$ in section 9. This 
is then identified with the coordinate of a string on the world sheet.
In the rest of the section, we derive the general form of the action
for $v$. It  turns out to be the action for a free field in two
dimensions, with, however, a non-trivial dispersion. In the next
section, we work this action out in detail as a function of a
momentum variable $k$ conjugate to $\sigma$, and also as a function
of the integer $n$, with $k=2\,\pi\,n$. This discretization is due to
compactification of $\sigma$ on circle of unit perimeter. The square
of the mass of the excitations on the string trajectory consists of
three terms: The leading term is linear in $n$, and then there is
a non-leading 
logarithmic correction and a constant term. We end the section with
some concluding remarks. We argue that what we have is an asymptotic
expansion of the string trajectory in the variable $n$. Within the
context of our ansatz, the terms calculated are exact, and the terms
we have dropped vanish as $n\rightarrow \infty$. The main conclusion
of the paper is that, in the variational approximation, the $QCD\,3$
string trajectory is asymptotically linear, with however, low energy
corrections.

In the final section, we  summarize our results and discuss
directions for future research.

\section{The World Sheet Picture}
\vskip 9pt

The planar graphs of the free part of $QCD\,3$ are the same as in the massless
scalar $\phi^{3}$ theory. They
can be represented on a world sheet
parametrized by the $\tau=x^{+}$ and $\sigma=p^{+}$ as a collection of
horizontal solid lines (Fig.1), where the n'th line carries the one
dimensional transverse momentum $q_{n}$.
\begin{figure}[t]
\centerline{\epsfig{file=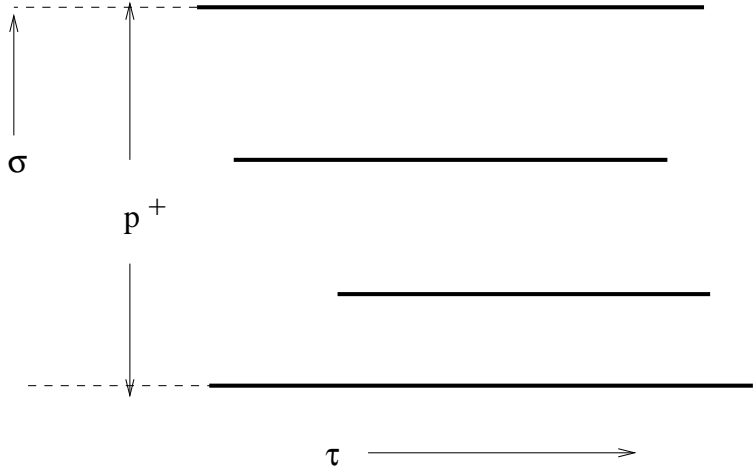, width=10cm}}
\caption{A Typical Graph} 
\end{figure}
Two adjacent solid lines labeled by n and n+1 correspond to the light
cone propagator
\be
\Delta(p_n)=\frac{\theta(\tau)}{2 p^{+}}\,\exp\left(-i
\tau\,\frac{p_{n}^{2} }{2\,p^{+}}\right),
\ee
where $p_n= q_{n+1}- q_{n}$ is the transverse momentum  and
$$
p^{+}_{n}=\sigma_{n+1} -\sigma_{n},
$$
is the light cone momentum flowing through the propagator.

 In the interacting theory, in addition to the 
propagators, there are three and four point vertices. The two three
point vertices are pictured in Fig.2. When lines 1 and 2 merge to form
the line 3, the associated vertex factor is given by
\be
V(1+2\rightarrow 3)=
\left(\frac{\sigma_{2} -\sigma_{1}}{\sigma_{3} -\sigma_{2}}
+\frac{\sigma_{3}-\sigma_{2}}{\sigma_{3}-\sigma_{1}}\right)\,p_{2}
-\left(\frac{\sigma_{2} -\sigma_{1}}{\sigma_{3} -\sigma_{1}} +
\frac{\sigma_{3} -\sigma_{2}}{\sigma_{2} -\sigma_{1}}\right)\,
p_{1}.
\ee

The vertex factor $V(3\rightarrow 1+2)$, for line 1 splitting into
lines 2 and 3, is given by the conjugate expression. We will not write
down the four point vertex since it will not be needed in the present work.

\begin{figure}[t]
\centerline{\epsfig{file=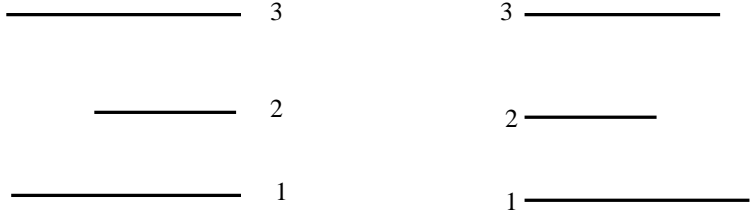, width=10cm}}
\caption{Interaction Vertices}
\end{figure}

\vskip 9pt

\section{The World Sheet Field Theory}

\vskip 9pt

The light cone graphs described above are generated by a world sheet
field theory. We introduce a complex scalar field $\phi(\tau,\sigma,
q)$ and its conjugate $\phi^{\dagger}$, which at time $\tau$,
annihilate (create) a solid line with coordinate $\sigma$, carrying
momentum $q$. They satisfy the usual commutation relations:
\be
[\phi(\tau,\sigma, q), \phi^{\dagger}(\tau,\sigma',q')]=\delta(\sigma-
\sigma')\,\delta(q-q').
\ee
The vacuum, annihilated by the $\phi$'s, represents the empty world
sheet.
For later use, it is also convenient to define the composite operator
$\rho$ which represents the density of the solid lines:
\be
\rho^{2}(\tau,\sigma)=\int
dq\,\phi^{\dagger}(\tau,\sigma,q)\,\phi(\tau,\sigma,q).
\ee

The free Hamiltonian consists of a bunch of solid lines, representing
free propagators. An important restriction is that propagators are
assigned to adjacent solid lines, and not to the non-adjacent ones. To
enforce this constraint, we need to define the projection operator 
$\mathcal{E}(\sigma_{i},\sigma_{j})$. The projection operator
 is defined by the equations
$$
\mathcal{E}(\sigma_{i},\sigma_{j})|s\rangle =0
$$
if $\sigma_{j}\leq \sigma_{i}$.
$$
\mathcal{E}(\sigma_{i},\sigma_{j})|s\rangle =0
$$
if $\sigma_{j}>\sigma_{i}$ \underline{and} there are solid lines
between $\sigma=\sigma_{i}$ and $\sigma=\sigma_{j}$.
$$
\mathcal{E}(\sigma_{i},\sigma_{j})|s\rangle=|s\rangle
$$
if  $\sigma_{j}>\sigma_{i}$ \underline{and} there are no solid lines
between  $\sigma=\sigma_{i}$ and $\sigma=\sigma_{j}$. These equations
are all that is needed to compute the matrix elements
$\langle s|\mathcal{E}(\sigma_{i},\sigma_{j})|s\rangle $ and derive
equations (6.5) and (6.9).
Also, using the  properties of the projection operator described above,
 the free Hamiltonian
can be written as
\bq
H_{0}&=&\frac{1}{2} \int d\sigma \int d\sigma' \int dq \int dq'\,\frac{
\mathcal{E}(\sigma,\sigma')}{\sigma' -\sigma}\, (q-q')^{2}\nonumber\\
&\times& \phi^{\dagger}(\sigma,q)\phi(\sigma,q)\,\phi^{\dagger}(\sigma',q')
\phi(\sigma',q')\nonumber\\
&+&\int d\sigma\,\lambda(\sigma) \left(\int dq\,\phi^{\dagger}(\sigma,q)
\phi(\sigma,q) -\rho^{2}
(\sigma)\right),
\eq
where $\lambda$ is a Lagrange multiplier.

The interaction Hamiltonian, which reproduces vertex factors of (2.2), is
given by
\bq
H_{I}&=& ig\,
\int d\sigma_{1} \int d\sigma_{2} \int d\sigma_{3}\,
\theta(\sigma_{2}-\sigma_{1})\,\theta(\sigma_{3}-\sigma_{2})\,
\frac{\mathcal{E}(\sigma_{1},\sigma_{3})}{\sqrt{(\sigma_{2}-\sigma_{1})\,
(\sigma_{3}-\sigma_{2})\,(\sigma_{3}-\sigma_{1})}}\nonumber\\
&\times&\left(1+\frac{\sigma_{3}-\sigma_{2}}{\sigma_{2}-\sigma_{1}}+
\frac{\sigma_{2}-\sigma_{1}}{\sigma_{3}-\sigma_{2}}\right)\,
\rho^{2}(\sigma_{1})
\,\rho^{2}(\sigma_{3})\,\int dq_{2}\,q_{2}\,\phi(\sigma_{2},q_{2})+H.C.
\eq
The $\theta$ functions order the $\sigma$ integrations so that
$\sigma_{1}<\sigma_{2}<\sigma_{3}$. The total Hamiltonian
\be
H=H_{0}+H_{I},
\ee
as well as the commutation relations (3.1), follow from the action
\be
S=\int d\tau\,\left(i\,\int d\sigma \int dq\,\phi^{\dagger}\partial_{\tau}
\phi\,-H(\tau)\right).
\ee

An important feature of this action is its symmetries.
It is invariant 
under the light cone subgroup of Lorentz transformations, and also
under translations of the transverse momentum,
\be
\phi(\tau,\sigma,q)\rightarrow \phi(\tau,\sigma, q +r),
\ee  
by a constant $r$, as well as translations in $\sigma$ and $\tau$
coordinates.
 Among the lightcone symmetries, the boost along
the special direction 1 is of special importance. Under this
transformation, parametrized by $u$, the fields transform as
\be
\phi(\tau,\sigma,q)\rightarrow\sqrt{u}\,\phi(u\tau, u\sigma, q),\,\,
\lambda(\tau,\sigma)\rightarrow u\,\lambda(u\tau,u\sigma),\,\,
p^{+}\rightarrow\frac{1}{u}\,p^{+}.
\ee
To simplify the algebra, we take advantage of this invariance and set,
\be
p^{+}=1,
\ee
by taking $u=p^{+}$. The correct $p^{+}$ dependence can always be
restored at the end of a calculation.

Another important symmetry is
\be
\phi(\tau,\sigma,q)\rightarrow -\phi(\tau, \sigma,-q),\,\,
\phi^{\dagger}(\tau,\sigma,q)\rightarrow -\phi^{\dagger}(\tau,
\sigma,-q).
\ee

These symmetries allow us to simplify the search for the ground state.
We follow the common practice and assume that the ground state
configuration is invariant under these symmeries.

\vskip 9pt
 
\section{The Setup For The Variational Calculation}

\vskip 9pt
 
In the standard variational approach, the approximate ground state
energy and the wave function is computed by sandwiching the
Hamiltonian between suitably chosen trial states and minimizing the
energy with respect to the variational parameters. In our case, an
arbitrary state is generated by applying a product of $\phi^{\dagger}$'s 
at various values of $\sigma$'s and $q$'s but at a fixed value of $\tau$
to the vacuum. In this section, we will introduce the trial state
we will use and carry out part of the variational calculation. The
motivation for the choice of this state was explained in [1].The
variational state is given by
\be
|s\rangle=\sum_{n=1}^{\infty}|n,\sigma=1\rangle,
\ee
where the states on the right hand side of this equation are defined
by the recursion relation
\be
|n+1,\sigma\rangle=K(\sigma)\,\int_{0}^{\sigma}
d\sigma'\,f(\sigma-\sigma')
|n,\sigma'\rangle,
\ee
and the initial condition
\be
|n=0,\sigma\rangle=|0\rangle.
\ee
Here $n$ is a positive integer and $\sigma$ ranges from $0$ to
$p^{+}=1$. The correlation function $f$ will be specified later. 
We note that the sum over $n$ starts at $n=1$, so that the empty
world sheet is eliminated.
 
We complete the specification of the trial state
by taking  for $K$
\be
K(\sigma)=\int dq\,A(\sigma,q)\,\phi^{\dagger}(\sigma,q).
\ee

We note that in this ansatz, the dependence on $q$ and $\sigma$
factorizes. It is then easy to show that the contribution of the four
point vertex vanishes. This is, of course, a feature of this particular
ansatz and is not true in general.

We now have to compute the normalized expectation value of the
Hamiltonian,
\be
\langle H \rangle\equiv N^{-1}\, \langle s|H|s \rangle,
\ee
as a function of  the variational parameters
of the problem, and  solve the corresponding variational
equations. $N$ is the normalization constant given by
$$
N=\langle s|s \rangle.
$$

 We take the solutions to these equations,
 $A(q)$, $\lmz$ and $\rhz$ to
be independent of
$\sigma$. This is because the ground state wavefunction is expected to
be invariant under the symmetries of the problem, in this case,
translation invariance in $\sigma$. We will discuss invariance under
translations of $q$ later on. From now on, we will use the notation
$$
\langle O \rangle=N^{-1}\,\langle s|O|s \rangle
$$
for the normalized expectation value of any operator $O$.

 In the next section, we will  solve the variational equation 
  for $A$, and leave the
rest to the subsequent sections.

\vskip 9pt

\section{The Variational Equation For A}

\vskip 9pt

By sandwiching $H$ between the states $|s \rangle$, it is easy to show
that the normalized expectation values of the various terms of the
Hamiltonian are of the form
\bq
\langle H_{0} \rangle&=&Z_{0}\,\int dq\,q^{2}\, |A(q)|^{2},\nonumber\\
\langle H_{I} \rangle&=&i g\,Z_{I}\,\int dq\,q\,
\left(A(q) -A^{\star}(q)\right),
\eq
and,
\be
\langle H \rangle=\langle H_{0} \rangle +\langle H_{I} \rangle+
\lmz\,\left(\int dq\,
|A(q)|^{2} -\rhz^{2}\right).
\ee

Here,
\bq
\lambda_{0}&=&\langle \lambda \rangle,\,\,\rho_{0}=
\langle \rho \rangle,\nonumber\\
Z_{0}&=&\int_{0}^{1} d\sigma\,Z(\sigma),\nonumber\\
Z(\sigma'-\sigma)&=&\langle \frac{ 
\mathcal{E}(\sigma, \sigma')}{\sigma' -\sigma}
\rangle.
\eq
  Actually, $\rhz$ is a redundant parameter; it can be
absorbed into definition of $f$, so we will set $\rhz=1$ from now on.
We will give the expression for $Z_{I}$ later.

The variational equation,
\be
\frac{\delta\langle H \rangle}{\delta A^{\star}(q)}
=Z_{0}\,q^{2}\,A(q)+\lmz\,A(q)-i g\,Z_{I}\,q=0,
\ee
has the solution
\be
A(q)=i g\,Z_{I}\,\frac{q}{Z_{0}\,q^{2}+\lmz}.
\ee
 From
\be
\int dq\,|A(q)|^{2}=\rhz^{2}=1
\ee
it follows that
\be
\lmz=\frac{\pi^{2}}{4}\,g^{4}\,Z_{I}^{4}\,Z_{0}^{-3}.
\ee
Defining a mass parameter by
\be
\mzs=\frac{\lmz}{Z_{0}}=\frac{\pi^{2}}{4}\,
\left(\frac{g\,Z_{I}}{Z_{0}}\right)^{4},
\ee
$A(q)$ can be rewritten as
\be
A(q)= ig\,\frac{Z_{I}}{Z_{0}}\,\frac{q}{q^{2}+\mzs}.
\ee
 
If we now try to compute $\langle H \rangle$ using (5.5),
 we find that the integral
over $q$ is linearly divergent. This is due to the translation invariance
in $q$ and the assignment of two transverse momenta with each internal
line. The momentum flowing through the line is then the difference of
these auxiliary momenta (eq.(2.1)). 
The energy is then proportional to the volume in
 momentum space, so the finite quantity is the
energy density.
 If we put the system in a  one dimensional box of
size $L$, 
  the energy density $E$ is given by
\be
E=\frac{\langle H \rangle}{L}\rightarrow -g^{2}\,\frac{Z^{2}_{I}}{Z_{0}}.
\ee

Our next task is to compute the constants $N$, 
$Z$ and $Z_{I}$.  We do this in the next
section by solving the recursion relation (4.2).

\vskip 9pt

\section{Solution Of The Recursion Relation For The Variational States}

\vskip 9pt

We start with the definitions
 $$
N(n,\sigma)=\langle n,\sigma|n,\sigma \rangle,
$$
and
$$
N(\sigma)=\sum_{n=1}^{\infty}N(n,\sigma).
$$
The normalization constant $N$ for the state $|s\rangle$ is then given by
\be
N= N(\sigma=1)=
\langle s|s \rangle.
\ee

The recursion relation (4.2) for the auxiliary states can be
rewritten as
\be
|n+1,\sigma \rangle =\int dq\,\int_{0}^{\sigma} d\sigma'
\,f(\sigma-\sigma')\,\eyz\,\phi^{\dagger}(\sigma,q)
|n,\sigma' \rangle,
\ee
and the corresponding recursion relation for $N(n,\sigma)$ is
\be
N(n+1,\sigma)=\int_{0}^{\sigma} d\sigma'\,
f^{2}(\sigma -\sigma')\,N(n,\sigma').
\ee
By  Fourier  transforming in the variable $\sigma$,
this is reduced to an algebraic equation, which is easily solved.
The result can be written as
\be
N(\sigma)=\int dk\,e^{i k\sigma}\,\frac{1}{1- 2\pi\,F(k)},
\ee
where,
$$
2\pi\,F(k)=\int_{0}^{\infty} d\sigma\,e^{-i k\sigma}\,f^{2}(\sigma).
$$
It is convenient to define $f(\sigma)$ so that it vanishes for
$\sigma<0$. This enables one to extend the Fourier integral as in the
above equation to all values of $\sigma$. We also note that $F(k)$ is
analytic for $Im(k)<0$ and vanishes as $Im(k)\rightarrow
-\infty$. This property of the Fourier transforms of  functions that
vanish on the half of the real line will be useful later on.

Now  consider $Z_{0}$ (eq.(5.3)). Defining
$$
\zbz=N\,Z_{0},
$$
it can be written as an infinite series:
\bq
\zbz&=&\int dk\,e^{i k}\,2\pi\,F_{1}(k)\,\sum_{n=0}^{\infty}\,
(n+1)\,(2\pi\,F(k))^{n}\nonumber\\
&=&\int dk\,e^{i k}\,\frac{2\pi\,F_{1}(k)}{(1 -2\pi\,F(k))^{2}},
\eq
where,
\be
2\pi\,F_{1}(k)=\int_{0}^{\infty} d\sigma\,e^{i k \sigma}\,
\frac{f^{2}(\sigma)}{\sigma}.
\ee
In this equation, the factor of $n+1$ counts the number of
distinct insertions of $\mathcal{E}(\sigma -\sigma')$ in the
$n'th$ term of the sum. Alternative expressions for $Z(\sigma)$ and
$\zbz$ are
\bq
Z(\sigma)&=&\frac{\cnb(\sigma)}{N}\,\frac{f^{2}(1 -\sigma)}
{1-\sigma}=\frac{1}{N}\,\int d k\,e^{i\,k\,\sigma}\,\frac{F_{1}(k)}
{\left(1-2\,\pi\,F(k)\right)^{2}},\nonumber\\
\zbz&=&\int_{0}^{1}
 d\sigma\,\cnb(1-\sigma)\,\frac{f^{2}(\sigma)}{\sigma},
\nonumber\\
\cnb(\sigma)&=&\int dk\,e^{i k\sigma}\,\frac{1}{(1-
 2\pi\,F(k))^{2}}.
\eq

Next, we define the matrix elements of the interaction Hamiltonian
(2.2) between the variational states by
\be
\langle s|H_{I}|s \rangle=
\sum_{n=1}^{\infty}\,\langle n,\sigma=1|H_{I}|n+1,\sigma=1 \rangle+H.C.
=Z_{I}=N\,\zbi.
\ee
A straightforward calculation gives
\bq
\zbi&=&N\,Z_{I}=\int_{0}^{1} d\sigma \int_{0}^{\sigma}d\sigma'\,
\cnb(1-\sigma)\,\left(\sigma\,\sigma'\,(\sigma-\sigma')\right)^{-1/2}
\nonumber\\
&\times&\left(1+\frac{\sigma-\sigma'}{\sigma'}+\frac{\sigma'}{\sigma
-\sigma'}\right)\,f(\sigma)\,f(\sigma')\,f(\sigma-\sigma').
\eq

 So far, we have not specified the function $f$, which is a part of the
trial wave function, and so it should be determined
  by minimizing the ground state energy (5.10):
\be
\frac{\delta E}{\delta f(\sigma)}\rightarrow \frac{\delta}{\delta f(\sigma)}
\left(\frac{Z_{I}^{2}}{Z_{0}}\right)=0.
\ee
This equation can be written in two more convenient equivalent forms:
\bq
0&=&
\frac{2}{Z_{I}}\,\frac{\delta Z_{I}}{\delta f(\sigma)} -\frac{1}{Z_{0}}\,
\frac{\delta Z_{0}}{\delta f(\sigma)},\nonumber\\
0&=&\frac{2}{\zbi}\,\frac{\delta \zbi}{\delta f(\sigma)}- 
\frac{1}{\zbz}\,\frac{\delta \zbz}{\delta f(\sigma)}-
\frac{1}{N}\,\frac{\delta N}{\delta f(\sigma)}.
\eq

This is the fundamental equation  for the variational function
$f$ corresponding to the ground state of the model. It is a
complicated non-linear equation, which at first sight looks intractable.
However, in the next section, we introduce an ansatz which enables us
to solve it by an iterative procedure.

\vskip 9pt

\section{The Variational Ansatz}

\vskip 9pt

The iterative procedure we are proposing is based on an expansion
of $f(\sigma)$
in powers of $\sigma$ around $\sigma=0$:
\be
f(\sigma)=\sum_{n=0}^{\infty}\beta_{n}\,\sigma^{\alpha_{n}}.
\ee
The constants $\alpha_{n}$ and $\beta_{n}$ are real numbers, and the
$\alpha$'s form an increasing sequence, with
$$
\alpha_{n+1} >\alpha_{n},
$$
and therefore, the terms with the most singular $\sigma$ dependence 
are those with the smallest values of $n$.
 These then dominate the
asymptotic limit $k\rightarrow\infty$ in the Fourier conjugate
variable $k$. This correspondence will later be very useful
in determining the asymptotic limits of the string
trajectories.

In this section, we are going to  compute $N(\sigma)$,
$\zbz$ and $\zbi$ as a series in terms of the expansion (7.1). 
 To get started, let us first consider the contribution of the first
term, $n=0$, 
$$
f(\sigma)= \baz\,\sigma^{\aaz},
$$
to $N$:
\be
2\pi\,F(k)= \int_{0}^{\infty} d\sigma\,\baz^{2}\,
\sigma^{2\aaz}\,e^{-i k \sigma} =\baz^{2}\,\Gamma(1+2\aaz)\,
(i k+\epsilon)^{-1-2\aaz} ,
\ee
and substituting this in eq.(6.4) gives $N$ as a Fourier transform.
To evaluate this integral, we first convert it into a Laplace
trasform over a real exponential. Noticing that the function
$(i k)^{-1-2\aaz}$ has branch cut on the positive imaginary axis,
we distort the contour integration in $k$ to wrap it around this
cut. The values of this function above and below the cut
are given by
\be
(i k\pm \epsilon))^{s}\rightarrow
|p|^{s}\,\exp(\pm i\pi s),
\ee
where $p=-i k$ and for convenience, we have also defined
$$
 s=1+2 \aaz.
$$
 Putting all of this together, we have,
\bq
N(\sigma)
&=&i\int_{0}^{\infty} d p\,e^{-p\,\sigma}\,\Big(\frac{1}{1- \baz^{2}\,
\Gamma(s)\,p^{-s}\,\exp(-i\pi\,s)}\nonumber\\
&-&\frac{1}{1- \baz^{2}\,
\Gamma(s)\,p^{-s}\,\exp(i\pi\,s)}\Big).
\eq

Let us recall that this equation was obtained by summing a power series
in $F(k)$, which is a resummation of a perturbation expansion.
So long as the denominator in the expression for $N$ is expandable in
powers of $F(k)$, the perturbation results will be reproduced, and
nothing new or interesting will emerge. We propose to get out of this
difficulty by fixing 
the constants $s$ and $\baz$  by
\be
\aaz=- 1/2\rightarrow s=0,\,\,\,\baz^{2}\,\Gamma(s)=1,
\ee
so that the denominator vanishes for all $p$ and the  perturbation
expansion breaks down. Here we differ from [1], where $\aaz$ was
taken to be $1$. Apart from being non-perturbative, another
  advantage of the present choice is that
 it is the correct starting point of the iteration procedure that
solves the fundamental equation (6.10). Also, as we shall see, it
leads to asymptotically linear string trajectories.

There is, however,  another  problem with setting $s=0$ or $\aaz= -1/2$;
 several integrals we will encounter will be divergent. We will
regularize these divergences by analytic regularization, allowing the
constants $s$ and $\baz$ to be complex. Starting with $s$
positive and sufficiently large, when everything is convergent,
we analytically continue to negative values of $s$. The divergences
will then show up as a singularity at $s=0$. We will later see that
in all quantities of interest, this singularity cancels out, and the result
is finite. Therefore, we will set
\be
\baz^{2}=1/ \Gamma(s)\rightarrow s,\,\,\baz\rightarrow s^{1/2},
\ee
and take the limit
\be
s\rightarrow 0, (\aaz\rightarrow - 1/2)
\ee
approaching from positive $s$,
only after we have a finite expression. Since in this limit, the
denominators in the expression for $N$ in eq.(7.4) vanish,
 to get
a well defined result, we have to go to the next term in the series (7.1)
by letting
$$
f(\sigma)=\baz\,\sigma^{\aaz}+\tilde{f}(\sigma),
$$
and then taking the limit of (7.5), with the result,
\be
1- 2\pi\,F(k)\rightarrow - 2\pi\,\tilde{F}(k),
\ee
where,
\bq
2\pi\,\tilde{F}(k)&=& \int_{0}^{\infty} d\sigma\,e^{-i k\sigma}\,
\tilde{f}^{2}(\sigma),\nonumber\\
2\,\pi\,\tilde{F}_{1}(k)&=&\int_{0}^{\infty} d\sigma\,e^{-i k\sigma}\,
\frac{\tilde{f}^{2}(\sigma)}{\sigma},\nonumber\\
N(\sigma)&=&-\frac{1}{2\pi}\,\int dk\,e^{i k\sigma}\,
\frac{1}{\tilde{F}(k)},\nonumber\\
\cnb(\sigma)&=& \frac{1}{(2\pi)^{2}}\,
\int dk\,e^{i k\sigma}\,\frac{1}{(\tilde{F}(k))^{2}}.
\eq
Plugging in these results in the expression for $\zbz$, we have,
\be
\zbz=-\cnb'(1)+\int_{0}^{\infty} d\sigma\,\cnb(1-\sigma)\,\frac
 {\tilde{f}^{2}(\sigma)}{\sigma},
\ee             
where the slash on $\cnb$ indicates the derivative with respect to
its argument. An alternative expression for $\zbz$ is,
\be
\zbz= -\cnb'(1)+\frac{1}{2\,\pi}\,\int d k\,e^{ik}\,
\frac{\tilde{F}_{1}(k)}{\tilde{F}^{2}(k)}.
\ee

Next, we will compute $\zbi$, again in the limit $s\rightarrow 0$.
 It will turn out that $Z_{I}$ has a singularity
proportional to $s^{- 1/2}$
 in this limit. This singularity cancels between $Z_{I}$ and
$\delta Z_{I}/\delta f(\sigma)$, so the contribution to the variational
equation is finite, and depends only on the finite factors that
multiply this singularity.

To compute these finite factors, we define,
\bq
\zbi&=& \int_{0}^{1} d\sigma\,N(1-\sigma)\,\sigma^{-1/2}\,
f(\sigma)\,L(\sigma),\nonumber\\
L(\sigma)&=&\int_{0}^{\sigma} d\sigma'\,\left(1+\frac{\sigma'}
{\sigma -\sigma'}+\frac{\sigma -\sigma'}{\sigma'}\right)\,
(\sigma'\,(\sigma -\sigma'))^{-1/2}\,f(\sigma')\,f(\sigma -\sigma').
\nonumber\\
&&
\eq

$L$ consists of three terms:
$$
L=L_{1}+L_{2}+L_{3},
$$
where,
\bq
L_{1}(\sigma)&=&\baz^{2}\,\int_{0}^{\sigma}d\sigma'\,(\sigma')^
{\aaz -1/2}\,(\sigma -\sigma')^{\aaz - 1/2}\,
\Big(1+ \frac{\sigma'}{\sigma -\sigma'}+
\frac{\sigma -\sigma'}{\sigma'}\Big)\nonumber\\
&=&\baz^{2}\,\sigma^{2 \aaz}\,\Big(\frac{\Gamma^{2}(\aaz+ 1/2)}
{\Gamma(2 \aaz + 1)}+\frac{2\,\Gamma(\aaz+ 3/2)\,\Gamma(\aaz - 1/2)}
{\Gamma(2 \aaz + 1)}\Big)\nonumber\\
&\rightarrow &\frac{4\,\baz^{2}\,\sigma^{2 \aaz}}{2\aaz+1}
=4\,\sigma^{2 \aaz}.
\eq
Here, $L_{1}$ has no $s^{-1/2}$ factor, but this factor emerges upon
integration over $\sigma$ in eq.(7.12). $L_{2}$, defined by
\bq
L_{2}(\sigma)&=&2\baz\,\int_{0}^{\sigma} d\sigma'\,(\sigma')^{s/2 -1}
\,(\sigma-\sigma')^{-1/2}\,\tilde{f}(\sigma -\sigma')\nonumber\\
&\times&\left(1+\frac{\sigma -\sigma'}{\sigma'}+\frac{\sigma'}
  {\sigma-\sigma'}\right),
  \eq
has a factor of $s^{-1/2}$, which we calculate below. This singularity
comes from the integration near $\sigma'=0$, which diverges as
$s\rightarrow 0$. We will encounter divergences of this form later on,
which come from integrals of the general form
\be
I=\int_{0}^{c} dx\,x^{\gamma- n}\,G(x),
\ee
in the limit $\gamma\rightarrow 0$, where $n$ is a positive integer.
The pole term in $\gamma$ we are interested in, is isolated by
expanding $G$ in power series in $x$, with the result
\be
I\rightarrow \frac{1}{\gamma}\,\frac{1}{(n-1)!}\,\left(\frac{
  d^{n-1} G(x)}{dx^{n-1}}\right)_{x=0}.
\ee

Applying this result to $L_{2}$, we have,
\be
L_{2}(\sigma)\rightarrow 2\,s^{-1/2}\,\left(\sigma^{-1/2}\,
\tilde{f}(\sigma) - 2\,\sigma^{1/2}\,\tilde{f}'(\sigma)\right).
\ee
The remaining term $L_{3}$ has no singularity. Putting all of this
together gives
\bq
s^{1/2}\,\zbi&\rightarrow& -\frac{8}{3}\,\cnb'(1)+ \int_{0}^{1}
d\sigma\,\cnb(1-\sigma)\,
\left(\frac{2 \tilde{f}^{2}(\sigma)}{\sigma} -4\,\tilde{f}(\sigma)
\,\tilde{f}'(\sigma)\right).\nonumber\\
&&
\eq

\vskip 9pt

\section{Iterative Solutions Of The Variational Equation}

\vskip 9pt

In this section, we are going to solve the variational equation,
using an ansatz of the type described in section 7. The specific
form of the ansatz is,
\be
\tilde{f}(\sigma)= \sum_{n=1}^{\infty}\,\beta_{n}\,\sigma^{\aao+n-1}.
\ee
Here, $\beta_{n}$ and $\aao$ are the variational parameters
to be determined; we get
an infinite number of equations
for them by setting the variation of $E$
with respect to each parameter equal to zero. These equations can then
be solved iteratively. In
this paper, we will only consider a more modest problem, where the
series in eq.(8.1) is truncated at the second term. We write it in the
form
\be
\tilde{f}(\sigma)= \beta_{1}\,(1+x\,\sigma)\,\sigma^{\aao},
\ee
where $x=\beta_{2}/\beta_{1}$. Varying with respect to $x$,
 we have,
\be
\frac{2}{\zbi}\,\frac{\partial \zbi}{\partial x} -\frac{1}{\zbz}
\,\frac{\partial \zbz}{\partial x}- \frac{1}{N}\,\frac{\partial N}
{\partial x}=0.
 \ee

 Anticipating the results to be derived, it turns out that
 $\aao\rightarrow 0$, and $x$ has several possible values including $x=0$
 , and $\bao$ is arbirary. This is why we have not written the equation
 with respect to $\bao$. Substituting the ansatz (8.2) in the equations
 (7.9),
 \bq
 2\,\pi\,\tilde{F}(k)&\rightarrow&\beta_{1}^{2}\,(ik)^{-1-2 \aao}\,
 \left(1+ \frac{2 x}{ik}+\frac{2 x^{2}}{(i k)^{2}}\right)\nonumber\\
 2\,\pi\,\tilde{F}_{1}(k)&\rightarrow&
 \beta_{1}^{2}\,(i k)^{-2\,\aao}\,\left(\frac{1}
 {2 \aao}+ \frac{2\,x}{(i k)}+\frac{x^{2}}{(i k)^{2}}\right).
 \eq
 Here, to simplify the algebra, we will keep only the leading
 terms as $\aao\rightarrow 0$. For example, in the expression for
 $\tilde{F}_{1}$, we will keep the term proportional to $1/\aao$ and
 drop finite terms. Of course, in the end, we will verify that
 this limit solves the variational equation.

 Substituting these expressions in eqs. (6.4, 6.5),
 \bq
 N(\sigma)&=&-\frac{4\pi\,\aao}{\bao^{2}}\,\int_{0}^{\infty}
 dp\,e^{-p\,\sigma}\,\frac{p}{1+\frac{2\,x}{p}+\frac{2\,x^{2}}
   {p^{2}}}+ P_{1}(\sigma),\nonumber\\
% \cnb(\sigma)&=&-\frac{8\,\pi\,\aao}{\bao^{4}}\,
% \int_{0}^{\infty}dp\,e^{-p\,\sigma}\,\frac{p^2}
% {\left(1+\frac{2\,x}{p}+\frac{2\,x^{2}}{p^{2}}\right)^{2}},
 %\nonumber\\
 \zbz&=&-\frac{2\,\pi}{\bao^{2}}\,  
 \int_{0}^{\infty}dp\,e^{-p}\,\frac{p^2}
 {\left(1+\frac{2\,x}{p}+\frac{2\,x^{2}}{p^{2}}\right)^{2}}
 + P_{2}.
\eq
Again, we have simplified by dropping higher order terms 
in $\aao$. The terms $P_{1,2}$ in this equation have the following source:
As one distorts the contour of integration from real $k$ to positive
imaginary $k$,one encounters poles at points where the denominator
vanishes. For $x<0$, these poles are in the lower half plane and they
do not contribute. Therefore, $P_{1,2}=0$. For $x>0$, the two poles are located
at
$$
i k=y_{\pm}=x\,(-1\pm i),
$$
and $P_{1,2}$ are the sum of the residues at these poles:
\bq
P_{1}(\sigma)
&=&\frac{e^{-\sigma\,x}}{\pi\,\bao^{2}}\,x^{2}\,\left(\cos(\sigma\,x)+
\sin(\sigma\,x)\right),\nonumber\\
P_{2}&=&\frac{e^{-x}}{\pi\,\aao\,\bao^{2}}\left(2\,x^{3}\,\cos(x)
+(3\,x^{3} -x^{4})\,\sin(x)\right).
\eq
It is clear that, in the limit $\aao\rightarrow 0$, the pole terms
$P_{1,2}$ dominate, so from now on, we will drop the integrals and
keep only the pole terms for $x>0$:
$$
N\rightarrow P_{1},\,\,\zbz\rightarrow P_{2}.
$$

For $x<0$, the pole terms are
absent and we are left with the integrals in (8.5).
We now show that $x=0$ is a solution to eq.(8.3). This solution has to
be defined as a limit approaching from the region $x<0$. We will see
that this specification is necessary, since there is a discontinuity
at $x=0$ from $P_{1,2}$, which contributes for $x>0$.
A simple calculation
shows that, approaching from $x<0$, with $P_{1,2}=0$,
\be
N|_{x=0}=\zbz|_{x=0}=-\frac{4\,\pi}{\bao^{2}},\,\,
\frac{\partial N}{\partial x}|_{x=0}=\frac{\partial \zbz}
{\partial x}|_{x=0}=\frac{8\,\pi}{\bao^{2}},
\ee
and the eq.(6.11) is clearly satisfied. Therefore,
\be
\tilde{f}=\bao\,  \sigma^{\aao}
\ee
is a solution in the limit $\aao\rightarrow 0$, with
$\bao$ arbitrary:
\be
\tilde{f}(\sigma)= \bao \,(\sigma)^{\aao}.
\ee

Although this is a mathematical solution for $x<0$, the presence of a
discontinuity at $x=0$ probably invalidates it as a solution to the
variational equation. We will therefore discard it and focus on the
the solutions for $x>0$,  which we will
investigate now. We first notice that the first and the last terms
on the right
in the equation (7.18) for $\zbi$ stay finite as $\aao\rightarrow 0$,
whereas the second term goes like $1/\aao$. Therefore, in this limit,
\be
s^{1/2}\,\zbi\rightarrow 2\,\int_{0}^{1} d\sigma\,\cnb(1-\sigma)\,
\frac{\tilde{f}(\sigma)^{2}}{\sigma}\rightarrow 2\,\zbz,
\ee
and  equation (8.3) simplifies:
\be
\frac{1}{\zbz}\,\frac{\partial \zbz}{\partial x}-
\frac{1}{N}\,\frac{\partial N}{\partial x}=0.
\ee
After a straightforwad calculation of the residues at the poles,
we have,
\bq
N=P_{1}&=&-\frac{x^{2}\,e^{-x}}{\pi\,\bao^{2}}\,\left(\cos(x)+\sin(x)\right),
\nonumber\\
\zbz=P_{2}&=&-\frac{4\,\pi\,e^{-x}}{\aao\,\bao^{2}}\,\left(2\,x^{3}\,
\cos(x)+(3\,x^{3} -x^{4})\,\sin(x)\right),
\eq
and  (8.3) then becomes,
\be
\frac{1}{x}-\frac{\cos(x)-\sin(x)}{\cos(x)+\sin(x)}
+\frac{(3 -x)\,\cos(x) -3\,\sin(x)}{2\,\cos(x)+(3-x)\,\sin(x)}
= 0.
\ee
Solving this equation numerically, the smallest solution is,
\be
x=x_{0}=1.41
\ee
There are other solutions with bigger values of $x$, which we have not
studied, hoping that the smallest value corresponds to the true ground
state with the minimum value of $E$.

To recapitulate, we have solved the  variational eq.(8.11) with
the trial function (8.2). The solution corresponds to the configuration
$$
i\phi_{i}=A(q),\,\,\,\phi_{r}=0,
$$
where $\phi_{i,r}$ are the real and imaginary parts of $\phi$, and $A$
is given by eq.(5.5), and $Z_{0}, Z_{I}$ by (6.5, 6.9). The corresponding
$m_{0}^{2}$ is,
\be
m_{0}^{2}=\frac{\pi^{2}}{4}\,\left(\frac{g\,Z_{I}}{Z_{0}}\right)^{4}
=4\,\pi^{2}\,g^{4}\,s^{-2}.
\ee

\vskip 9pt

\section{String Formation}

\vskip 9pt

In this section, we will consider  time dependent fluctuations 
around this static configuration. The particular fluctuation that leads to
string formation corresponds to
shifting the momentum $q$ by the fluctuating field 
$\vts$. We therefore start by letting
\be
i \phi_{i}\rightarrow A(q+v(\tau,\sigma)),
\ee
 In addition, $\phi_{r}$ is taken to be non-zero, and with $q$ again
 shifted by $\vts$:
\be
\phi_{r}\rightarrow \phi_{r}(\tau,\sigma,q+\vts).
\ee
We will see later that a non-zero $\phi_{r}$ is needed to have
the correct canonical quantization of the fields.

$A(q)$ originally broke translation invariance in $q$, since it was
localized around $q=0$. The introduction of the
 collective coordinate $v(\tau,\sigma)$ restores translation
invariance, since
$$
q\rightarrow q+r
$$
will be accompanied by
$$
v\rightarrow v-r.
$$
$v$ is then the Goldstone mode of the  symmetry generated by
translations in $q$.
It will also turn out to be the string coordinate. In this
paper, we will only consider fluctuations generated by $v$,
with all other parameters fixed at their ground state values.

If the ansatz given by (9.1) and (9.2) for $\phi_{i}$ and $\phi_{r}$ are
substituted in the kinetic energy term in the action (3.6), this term
 becomes,
\bq
K.E.&=& -2 \int d\tau \int d\sigma \int dq\,\phi_{r}(\tau,\sigma,
q+\vts)\,\partial_{\tau} \phi_{i}(\tau,\sigma,q+\vts)\nonumber\\
&\rightarrow& 2 i\, \int d\tau \int d\sigma \int
dq\,\phi_{r}(\tau,\sigma,q+\vts)\,\partial_{\tau}
A(q+\vts)\nonumber\\
&=& -2 g\,\frac{Z_{I}}{Z_{0}}\,
\int d\tau \int d\sigma \int dq\,\phi_{r}(\tau,\sigma,q)\,
\partial_{\tau}\vts\,
\partial_{q}\left(\frac{q}{q^{2}+\mzs}\right).\nonumber\\
&&
\eq
Here we have an action first order in the time $(\tau)$ variable,
with $\phi_{r}$ and $v$ as conjugate canonical variables.
This was the reason for keeping a non-zero $\phi_{r}$. Later,
$\phi_{r}$ will be eliminated using its equations of motion,
and the resulting action will depend only on $v$.

Next we consider the fluctuations of $\langle H \rangle$,
which all come from $\langle H_{0} \rangle$. As explained earlier,
the interaction term, which is linear in $\phi$, is eliminated
by shifting $\phi$ by $A$.
 We then make the replacement given by (9.1) and (9.2), and then
change the variable of integration from $q$ to $q- \vts$. The result is
\bq
\langle H \rangle &\rightarrow& \frac{1}{2} \int d\sigma \int d\sigma'
\int d q \int d q'\,Z(\sigma' -\sigma)\,
\left(q -q'+v(\sigma')-
v(\sigma)\right)^{2}\nonumber\\
&\times& \phi^{\dagger} \phi(\sigma,q)\,\phi^{\dagger}
\phi(\sigma',q')
+ \int d\sigma\,\lambda(\sigma)\left(\int dq\,\phi^{\dagger}
\phi(\sigma,q) -1 \right).
\eq

Expanding in powers of $q$ and $q'$, terms linear in $q$ and $q'$
involve the integral 
$$
\int dq\,q\,\phi^{\dagger} \phi(\sigma,q)=0,
$$
which vanishes
because of the symmetry (3.10). We can therefore set,
\be
\langle H \rangle= \langle \tilde{H} \rangle +\langle H_{v} \rangle,
\ee
where $\langle \tilde{H} \rangle$ is $v$ independent and 
$\langle  H_{v} \rangle$ is quadratic in $v$:
\bq
 \langle H_{v} \rangle &=&\frac{1}{2} \int d\sigma \int d\sigma'
\int d q \int d q'\,Z(\sigma' -\sigma)
 \,(v(\sigma) - v(\sigma'))^{2} \,
\phi^{\dagger} \phi(\sigma,q)\,\phi^{\dagger}
\phi(\sigma',q')\nonumber\\
&=& \frac{1}{2}\, \int d\sigma \int d\sigma'\,
Z(\sigma' -\sigma)\,
(v(\sigma) - v(\sigma'))^{2} .
\eq

It is now convenient to go to momentum space by defining
\bq
v(\sigma)&=&\frac{1}{2\pi}\,\int d k\,e^{-i\,k\,\sigma}\,\tilde{v}(k),
\nonumber\\
Z(\sigma)&=&\int d k\,e^{i\,k\,\sigma}\,\tilde{Z}(k).
\eq
To simplify writing, we have suppressed the $\tau$
dependence of $\tilde{v}$ and $\tilde{Z}$. We remind the reader that
$v(\sigma)$ and $Z(\sigma)$ are defined to vanish for $\sigma<0$, and
therefore, $\tilde{v}(k)$ is analytic and bounded for $Im(k)>0$, and
$\tilde{Z}(k)$ is analytic and bounded for $Im(k)<0$. With these
definitions, eq.(9.6) becomes,
\be
\langle H_{v} \rangle =\int dk\,\left(\tilde{Z}(0)-\tilde{Z}(k)
\right)\,\tilde{v}(k)\,\tilde{v}(-k).
\ee
The computation of $\tilde{Z}(k)$ simplifies by noting that only terms
that are even under $k\rightarrow -k$ contribute:
$$
\tilde{Z}(k)\rightarrow \frac{1}{2}\,\left(\tilde{Z}(k)+\tilde{Z}(-k)
\right).
$$

Next, we have to compute $\langle \tilde{H} \rangle$ in the same
limit of the parameters. $\langle \tilde{H} \rangle$ is given by
(9.4), with $v=0$:
\bq
\langle \tilde{H} \rangle 
&=& \int d\sigma' \int d\sigma\,Z(\sigma' -\sigma)
 \, \int dq\,
q^{2} \, \phi^{\dagger} \phi(\sigma,q)\nonumber\\
&+&
\int d\sigma \, \lmz \,\left(\int dq \,\phi^{\dagger}
\phi(\sigma,q) \,-1\right)\nonumber\\
&=&Z_{0}\,\int_{0}^{1} d\sigma \int dq\,(q^{2}+ m_{0}^{2})\,
\phi_{r}^{2}(\sigma,q)-\lambda_{0},
\eq
where eqs.(5.3) and (5.8) have been used. Here, we have dropped a
quartic term in $\phi_{r}$. We will later argue that, in the limit
 $\aao\rightarrow 0$, this term vanishes.

The total action is the sum of (9.3),(9.6) and (9.9). 
The dependence on $\phi_{r}$ in this action can  be eliminated using its
equations of motion:
\be
\phi_{r}= -g\,\frac{Z_{I}}{Z_{0}^{2}\,(q^{2} +m_{0}^{2})}
\,\partial_{\tau}\vts\,\partial_{q}\left(\frac{q}{q^{2}+\mzs}\right),
\ee
and substituting  in (9.3), and making use of (8.10), we have, 
$$
K.E.= \int d\tau \int d\sigma\, \frac{7\,\pi\,g^{2}}
{32\,Z_{0}\,m_{0}^{5}\,s}\,\left(\partial_{\tau} v(\tau,\sigma)\right)^{2}.
$$

Finally, adding this to (9.8), the action in the momentum space is,
\bq
S&=&\int d\tau\,\int d k
\left(\frac{7\,g^{2}}{16\,Z_{0}\,s\,m_{0}^{5}}\,\partial_{\tau}
\tilde{v}(k)\,\partial_{\tau}\tilde{v}(-k)+
\left(\tilde{Z}(k)-\tilde{Z}(0)\right)\,\tilde{v}(k)\,
\tilde{v}(-k)\right).\nonumber\\.
&&
\eq

This action can be simplified by defining
\be
\tilde{v}(\tau,k)=\left(\frac{8\,Z_{0}\,s\,m_{0}^{5}}
{7\,g^{2}}\right)^{1/2}\,w(\tau,k),
\ee
with the result,
\bq
S&=&\int d\tau\,\int d k\left(\frac{1}{2}\,\partial_{\tau} w(\tau,
k)\,\partial_{\tau} w(\tau, -k)- \frac{1}{2}\,M^{4}(k)\,
w(\tau, k)\,w(\tau, -k)\right),\nonumber\\
M^{4}(k)&=&-\frac{16\,Z_{0}\,s\,m_{0}^{5}
  (\tilde{Z}(k)-\tilde{Z}(0))}{7\,g^{2}}.
\eq
In the next section, using this action,
we will determine the string trajectory in the
asymptotic limit of large $k$.

\vskip 9pt

\section{Corrections To The Linear Trajectory}

\vskip 9pt

The asymptotic limit of the string trajectory can be deduced from the
large $k$ limit of $\tilde{Z}(k)-\tilde{Z}(0)$, keeping only terms
even under $k\rightarrow -k$. The starting point is the equation

\bq
N\,\left(\tilde{Z}(k)-\tilde{Z}(0)\right)&=&\frac{1}{2\,\pi}\,
\int d\sigma\,\bar{N}(\sigma)\,\frac{\tilde{f}^{2}(1-\sigma)}
{1-\sigma}
(e^{-i\,k\,\sigma} -1)\nonumber\\
&=&\frac{1}{2\,\pi}\,\int d\sigma\,(e^{-i\,k\,\sigma} -1)\,
\int
dk_{1}\,\frac{e^{i\,k_{1}\,\sigma}}{(2\,\pi\,\tilde{F}(k_{1}))^{2}}\,
\int dk_{2}\,e^{i\,k_{2}\,(1-\sigma)}\,\tilde{F}_{1}(k_{2})\nonumber\\  
&=&\int d k_{1}\,e^{i\,k_{1}}\,\frac{\tilde{F}_{1}(k+k_{1})
  -\tilde{F}_{1}(k_{1})}{\left(2\,\pi\,\tilde{F}(k_{1})\right)^{2}}
\nonumber\\
&\rightarrow&\frac{1}{4\,\pi\,\aao\,\bao^{2}}\,\int d k_{1}\,
e^{i\,k_{1}}\,\frac{(i\,k_{1})^{6+4\,\aao}\,\left((i\,k+i\,k_{1})^{-2\,\aao}  
- (i\,k_{1})^{- 2\,\aao}\right)}{\left((i k_{1})^{2}+2\,x_{0}\,(i k_{1})
 +2\,x_{0}^{2}\right)^{2}},\nonumber\\ 
&&
\eq
which follows from eqs.(6.7) and (9.7).
  Since we are going to take the limit $\aao\rightarrow 0$, in the
  last step, we have kept only the leading term for $\tilde{F}_{1}$, which
  goes like $1/\aao$ (eq.(8.4)). Also, since $k=2\,\pi\,n$, we
  have set $e^{k}=1$.
%\eq
%\frac{1}{4\,\pi\,\aao\,\bao^{2}}\,\int dk_{1}\,e^{i\,k_{1}}\,
%\frac{(i\,k_{1})^{6+4\,\aao}\,\left((i\,k+i\,k_{1})^{-2\,\aao}
%  - (i\,k_{1})^{-2\,\aao}\right)}{(i\,k_{1}-y_{+})^{2}\,(i\,k_{1}-
 % y_{-})^{2}},
%\eq

As we did before, we now distort the contour of integration of
$k_{1}$ and wrap it around the branch cuts that go from $k_{1}=0$
to $k_{1}=i\,\infty$ or from $k_{1}=k$ to $k_{1}=k+i\,\infty$. We
denote this contribution by $\tilde{Z}_{c}(k)$. In addition, the
contour will cross two poles at $y_{\pm}=\xno\,(-1\pm i)$, and the
contribution from the residues will be denoted by $\tilde{P}(k)$.
The result is
\be
N\,\left(\tilde{Z}(k)-\tilde{Z}(0)\right)=\tilde{P}(k)+
N\,\left(\tilde{Z}_{c}(k)-\tilde{Z}_{c}(0)\right).
\ee

The contribution from the branch cuts can be computed as in section
7. We take first the limit $\aao\rightarrow 0$, and then the large $k$
asymptotic limit, keeping only the terms that do not vanish in this
limit:
\bq
&&\bao^{2}\,N\,\left(\tilde{Z}_{c}(k)-\tilde{Z}_{c}(0)\right)
\rightarrow\int_{0}^{\infty} dp\,e^{-p}\,\Big(\frac{(p-i\,k)^{6}}
    {\left((p-i\,k)^{2}+2\,\xno\,(p-i\,k)+2\,\xno^{2}\right)^{2}}\nonumber\\
   &-&\frac{p^{6}}{\left(p^{2}+2\,\xno\,p+2\,\xno^{2}\right)^{2}}\Big)
    \nonumber\\
    &\rightarrow&-k^{2}+4\,\xno^{3}\,\int_{0}^{\infty} dp\,e^{-p}\,
    \frac{2\,p^{3}+9\,\xno\,p^{2}+12\,\xno^{2}\,p+8\,\xno^{3}}
         {\left(p^{2}+2\,\xno\,p+2\,\xno^{2}\right)^{2}}.       
\eq

To evaluate the pole term $\tilde{P}(k)$, we first take the limit
$\aao\rightarrow 0$ in eq.(10.1) and symmetrize with respect to the sign
of $k$:
$$
(i\,k+i\,k_{1})^{-2\,\aao} -(i\,k_{1})^{-2\,\aao}\rightarrow
-\aao\,\ln\left(1-\frac{k^{2}}{k_{1}^{2}}\right).
$$
$\tilde{P}$ is then the sum of the residues at the poles $i\,k_{1}=
y_{\pm}$ in the expression
$$
-\frac{1}{4\,\pi\,\bao^{2}}\,\int d k_{1}\, e^{i\,k_{1}}\,\frac{
  (i\,k_{1})^{6}\,\ln\left(1-\frac{k^{2}}{k_{1}^{2}}\right)}
{(i\,k_{1}-y_{+})^{2}\,(i\,k_{1}- y_{-})^{2}},
$$
and, in the large $k$ limit, the result is,
\bq
\bao^{2}\,\tilde{P}(k)&\rightarrow& e^{-\xno}\,\xno^{3}\,\Big(
\left(2\,\cos(\xno) +(3 -\xno)\,\sin(\xno)\right)\,
\ln\left(\frac{k^{2}}{2\,\xno^{2}}\right)\nonumber\\
&+&\left(-1 +\frac{3\,\pi}{2} -\frac{\pi\,\xno}{2}\right)\,
\cos(\xno) -(1+\pi)\,\sin(\xno)\Big).
\eq

Adding up the cut and pole contributions from eqs.(10.3 and (10.4)),
and numerically evaluating at $\xno=1.41$, the result can be written as
\be
\bao^{2}\,\left(\tilde{Z}(k) -\tilde{Z}(0)\right)\rightarrow
- k^{2}+ 1.29\,\ln\left(\frac{k^{2}}{3.98}\right) +9.28.
\ee
Substituting in (9.13), we get a complicated equation for
$M^{4}$.  This  can then be simplified by defining a
physical mass term  $m$ by, and eliminating the dimensional parameter
$g$ in favor of the dimensional parameter $m$:
\be
\frac{16\,Z_{0}\,m_{0}^{5}\,s}{7\,g^{2}\,\bao^{2}}=
\frac{2^{9}\,\pi^{5}\,g^{8}}{7\,s^{4}\,\aao\,\bao^{2}}\,(Z_{0}\,\aao)
= m^{4},
\ee
where
$$
\aao\,Z_{0}= 4\,\pi^{2}\,\frac{2\,\xno\,\cos(\xno)+(3\,\xno -
  \xno^{2})\,\sin(\xno)}{\cos(\xno)+\sin(\xno)}= 91.7.
$$

As we let $s\rightarrow 0$ and $\aao \rightarrow 0$, we keep $m$ fixed
and finite, and define $g$ by eq.(10.6). With this definition, action
(9.13) becomes,
\bq
S&=&\int d\tau \int dk\,\Big(\frac{1}{2}\,\partial_{\tau} w(\tau,\,k)\,
\partial_{\tau} w(\tau,\,-k)\nonumber\\
&-& \frac{m^{4}}{2}\,\left(k^{2} -1.29\,
\ln\left(\frac{k^{2}}{3.98}\right) -9.28\right)\,w(\tau,\,k)\,
w(\tau,\,-k)\Big).
\eq

The spectrum of the string is determined by quantizing this action,
which is essentially the action for a two dimensional free field.
The square of the mass of a state on the trajectory is given by
\be
M^{2}=m^{2}\,\left(k^{2} -1.29\,\ln\left(\frac{k^{2}}{3.98}\right)
-9.28 \right)^{1/2}.
\ee
For plotting this function, we found it convenient to express
$k$ in terms of $n$ by letting $k=2\,\pi\,n$, and fixing
$m^{2}=1/(2\,\pi)$. Then, as a function of $n$,
\be
M^{2}=\left(n^{2}- 0.327\,\ln(9.919\,n^{2}) -0.235\right)^{1/2}.
\ee
This function is plotted in Fig.(3) from $n=1$ to $n=5$. It is
very close to a straight line.
\begin{figure}[t]
  \centerline{\epsfig{file=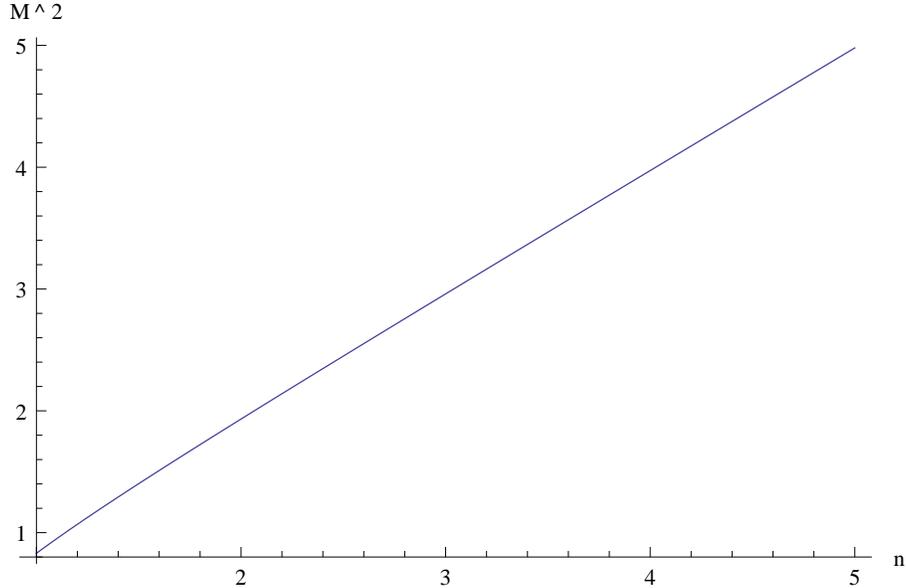, width=12 cm}}
  \caption{The String Trajectory}
\end{figure}

Now, we summarize the results of this paper with
some concluding comments:\\
a) The main result of this article is that, within our approximation
scheme,  $QCD 3$ has a spectrum represented
by  an asymptotically linear string trajectory. This, of course,
suggests that the theory is confining.\\
b) It would appear that there is a tachyon in the spectrum at $n=0$.
However, our truncated ansatz is only reliable for large
$n$. Consequently, we cannot say anything about the spectrum at
$n=0$.\\
c) We notice that $\bao$ has dropped out of the problem. This is
because the normalization of the variational state is proportional
to $1/\bao^{2}$, and physical quantities do not depend on this
normalization.\\
d) It is not hard to show that asymptotically vanishing contributions in
the large $n$ limit all come from terms proportional to
$\sigma^{1+\aao}$ and higher powers in the expansion (8.1), which we
have neglected. On the other hand, the terms which we have calculated,
which go like $n^{2}$, $\ln(n^{2})$ and a constant, receive no
contribution from the neglected terms in the expansion. Therefore,
they are exact within the context of the fundamental variational
calculation.\\
e) It seems like there is no parameter of expansion, since the
coupling constant is traded for a mass parameter $m$. Instead, the
expansion is an asymptotic one in $n$, and $1/n^{2}$ serves as an
expansion parameter. Although we will not do so here, higher order
terms in $1/n^{2}$ can be calculated by by solving the variational
equation for higher powers of $\sigma$ in the expansion of $f(\sigma)$.\\
f) In the equation for $\langle \tilde{H} \rangle$ (eq.9.9), we have
dropped quartic terms in $\phi_{r}$ and kept only the quadratic terms.
Using eq.(9.10), the quartic term can be expressed in terms of
$\partial_{\tau} w(\tau\,k)$. After some straightforward algebra, the
coefficient of this term turns out to be independent of $s$ and
proportional to $\aao^{3/2}$.  In the limit $\aao\rightarrow
0$, it vanishes and therefore it is consistent to drop it.

\vskip 9pt

\section{Discussion}

\vskip 9pt

In this article, we have extended the world sheet tratment of planar
$QCD\,3$ developed in an earlier work [1]. The main tool is again the
variational ansatz introduced there, but here we use a greatly
generalized version of the ansatz. It is then possible to solve
the variational equations in a systematic power series expansion. We
show that this expansion then leads to an asymptotic high energy
expansion of the string trajectory. We compute the first three terms
of this expansion, which are linear, logarithmic and constant in energy.
These are the main results of the present work.

There are several possible directions of future research suggested by
the present work. It should be possible to investigate $QCD\,3$ in
more detail by studying other possible fluctuations around the
background introduced here. Application of the variational approach
developed here to $QCD\,4$ also looks promising.

\vskip 9pt

\noindent{\bf Acknowledgement}

\vskip 9pt

This work was supported 
 by the Director, Office of Science,
 Office of High Energy  
 of the U.S. Department of Energy under Contract No.
DE-AC02-05CH11231.

\vskip 9pt

\noindent{\bf References}

\vskip 9pt

\begin{enumerate}

\item K.Bardakci, JHEP {\bf 07} (2019) 112.
\item M.B.Halpern, Phys.Rev. {\bf D 16}, (1977) 1798; 
 I.Bars and F.Green, Nucl.Phys. {\bf B 148}, (1979) 445;
J.Greensite, Nucl.Phys. {\bf B 158}, (1979) 469;
M.Bauer and D.Z.Freedman, Nucl.Phys. {\bf B 450}, (1995) 209;
O.Ganor and J.Sonnenschein, Int. J. Mod.Phys. {\bf A 11}, (1996) 5701;
D.Karabali and V.P.Nair, Nucl.Phys. {\bf B 464}, (1996) 135;
Pys.Lett. {\bf B 379},(1996) 141;
D.Karabali, Chanju Kim and V.P.Nair, Nucl.Phys. {\bf B 524}, (1998) 661.
\item K.Bardakci and C.B.Thorn, Nucl.Phys. {\bf B 626}, (2002) 286.
\item C.B.Thorn, Nucl.Phys. {\bf B 637}, (2002) 272;
S.Gudmundsson, C.B.Thorn and T.A.Tran, Nucl.Phys. {\bf B 649},(2003)
3-38.
\item C.B.Thorn and T.A.Tran, Nucl.Phys. {\bf B 677},(2004) 289.
\item G.'t Hooft, Nucl.Phys. {\bf B 72},(1974) 461.
\item K.Bardakci, JHEP {\bf 0810}, (2008) 056.
\item D.J.Gross and Y.Kitazawa, Nucl.Phys. {\bf B 206}, (1982) 440.

\end{enumerate}

\end{document}